\documentclass[showpacs,twocolumn,pre]{revtex4}
\usepackage{amssymb}
\usepackage{graphicx}
\usepackage{dcolumn}
\usepackage{bm}

\def\be{\begin{equation}}
\def\ee{\end{equation}}
\def\bea{\begin{eqnarray}}
\def\eea{\end{eqnarray}}
\def\ba{\begin{array}}
\def\ea{\end{array}}

\def\G{\Gamma}
\def\0{$\Gamma_0$}
\def\o{\omega}

\def\t{\theta}

\def\s{\sigma}

\begin{document}

\title{Alternative approach to all-angle-negative-refraction in two-dimensional
photonic crystals}
\author{Y. J. Huang}
\author{W. T. Lu}
\email{w.lu@neu.edu}
\author{S. Sridhar}
\email{s.sridhar@neu.edu}
\affiliation{Department of Physics and Electronic Materials Research Institute,
Northeastern University, Boston, Massachusetts 02115, USA}
\date{\today }

\begin{abstract}
We show that with an appropriate surface modification, a slab of photonic crystal
can be made to allow wave transmission within the photonic band gap. 
Furthermore, negative refraction and
all-angle-negative-refraction (AANR) can be achieved by this surface modification in frequency windows that 
were not realized before in
two-dimensional photonic crystals [C. Luo et al, Phys. Rev. B \textbf{65}, 201104 (2002)]. 
This approach to AANR leads to new applications in flat lens imaging.
Previous flat lens using photonic
crystals requires object-image distance $u+v$ less than or equal to the lens thickness $d$, $u+v \sim d$. 
Our approach can be used to design flat lens
with $u+v=\s d$ with $\s\gg1$, thus being able to image large and/or far away objects. Our
results are confirmed by FDTD simulations.
\end{abstract}

\maketitle

\section{Introduction}

Negative refraction (NR) was proposed theoretically a long time ago \cite%
{Veselago}. It was realized only recently in two classes of materials. One
type is the so-called meta-materials \cite{Shelby,Parazzoli,Shalaev},
consisting of wires and split-ring resonators. The second type is photonic
crystals \cite{Joannopoulos}, which have periodic permittivity and/or permeability.

In these artificial materials, there are up to now two mechanisms to achieve negative refraction (NR). 
One is to use the anti-parallelism between the
wave vector and the group velocity \cite{Notomi}. This can be realized in
isotropic metamaterials \cite{Shelby} and in the second or even higher band around the 
$\Gamma $ point in photonic crystals (PCs) \cite{Notomi}. In this case, the phase refractive
index $n_{p}$ is negative. 

The second mechanism is to use the anisotropy and the
concavity of the equi-frequency surface (EFS) such as the EFS around the
$M$ point in the first band of a square lattice PC. In this case though the phase refractive
index $n_{p}$ is positive, the group refractive index $n_{g}$ is negative 
\cite{Luo02}. In order to show negative lateral shift or all-angle-negative-refraction (AANR), one has to
orient the lattice such that the $\Gamma M$ direction is along the surface
normal. Negative refraction and flat lens imaging \cite{Pendry} using both mechanisms
have been observed in microwave \cite{Parimi04,Parimi03,Cubukcu,LuZ,Vodo05,Vodo06,Smith04} and
near infrared experiments \cite{Berrier}.

Recently we proposed a new mechanism for negative refraction using surface
grating \cite{Lu07}. This mechanism combines photonic band gap with surface
grating to achieve NR and AANR. Negative lateral shift and
AANR have been demonstrated
in a multilayered structure with surface grating. 

In this paper, we obtain new windows of AANR in two-dimensional (2D) 
PCs using this new mechanism of surface modification. 
We also show that flat lens made of
photonic crystal with surface grating can have $u+v=\s d$ \cite{Lu05} 
with $\s\gg 1$ while for the 
Veselago-Pendry flat lens \cite{Pendry} $\s=1$. Here $d$ is the thickness
of the flat lens, $u$ and $v$ are the distance from the lens to the object 
and the image, respectively.
Thus a flat lens can focus large and far away objects.

\section{New approach to AANR}

In a pioneering paper \cite{Luo02}, Luo et al showed that within certain
frequency windows in the first band of a PC, AANR can be
achieved. Specifically, within the first band, AANR is possible only along
the $\Gamma M$ direction for a square lattice PC. We will show that with
appropriate surface grating, NR and AANR is possible along the $\Gamma X$
direction in the first band of a square lattice PC.

\begin{figure}[htbp]
\vskip -4mm
\center{\includegraphics [angle=0,width=9cm]{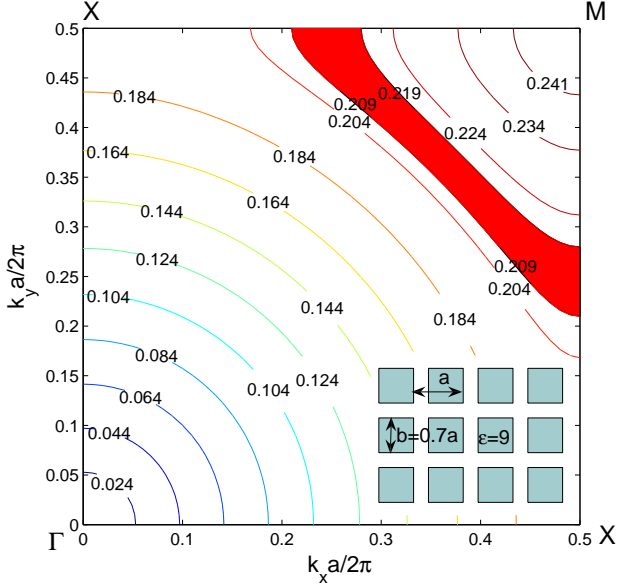}}
\vskip -4mm
\caption{(Color online) EFS of the TM modes of a square lattice PC.
The lattice is made of square rods of alumina ($\varepsilon=9$) with filling ratio $0.49$ (see insert).
The shading area is the window of AANR for the PC with surface grating $a_s=2a$ (see Fig. \ref{fig3}). 
A slab of this PC is oriented such that the surface normal is along the $\Gamma X$ direction.}
\label{fig1}
\end{figure}

In the main text of this paper, we only consider the transverse magnetic (TM) modes 
of a square lattice PC of square rods. Square lattice of circular rods or 
even rhombus rods can be treated similarly.
The generalization to transverse electric (TE) modes and 
lattice structures other than square lattice is straightforward. 
As a specific example, we consider a square lattice of square
rods with size $b/a=0.7$, thus a filling ratio of 0.49. 
The EFS is of this
PC is calculated using the plane-wave expansion \cite{Joannopoulos} with
5041 plane waves. The EFS of the first band is shown in Fig. \ref{fig1}. 
The frequencies at the $X$ point and the $M$ point are $\omega _{X}=0.1943\times 2\pi c/a$ 
and $\omega_{M}=0.2446\times 2\pi c/a$, respectively.  

\begin{figure}[htbp]
\vskip -2mm
\center{\includegraphics [angle=0,width=5.2cm]{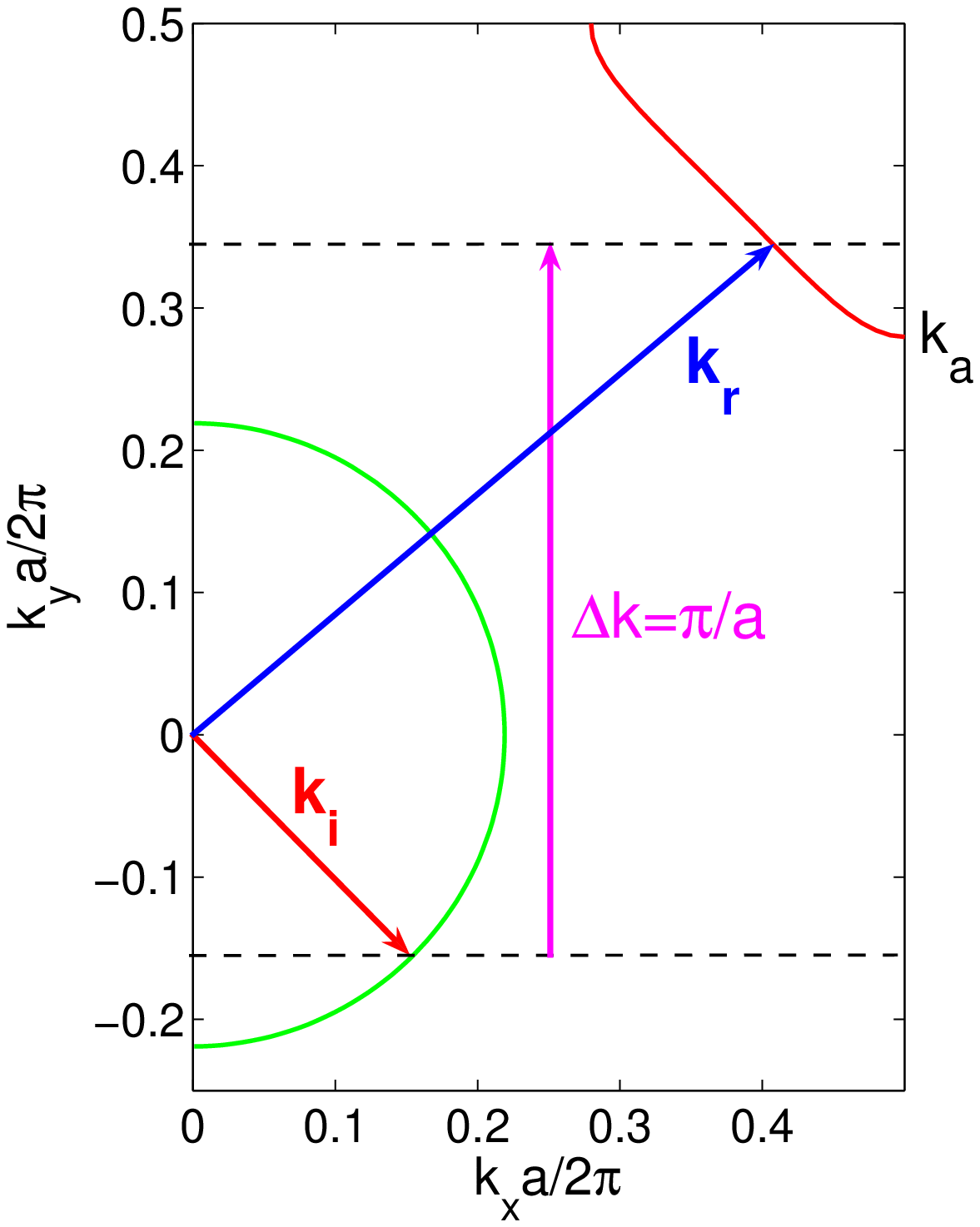}
\hskip -5mm
\includegraphics [angle=0,width=3.7cm]{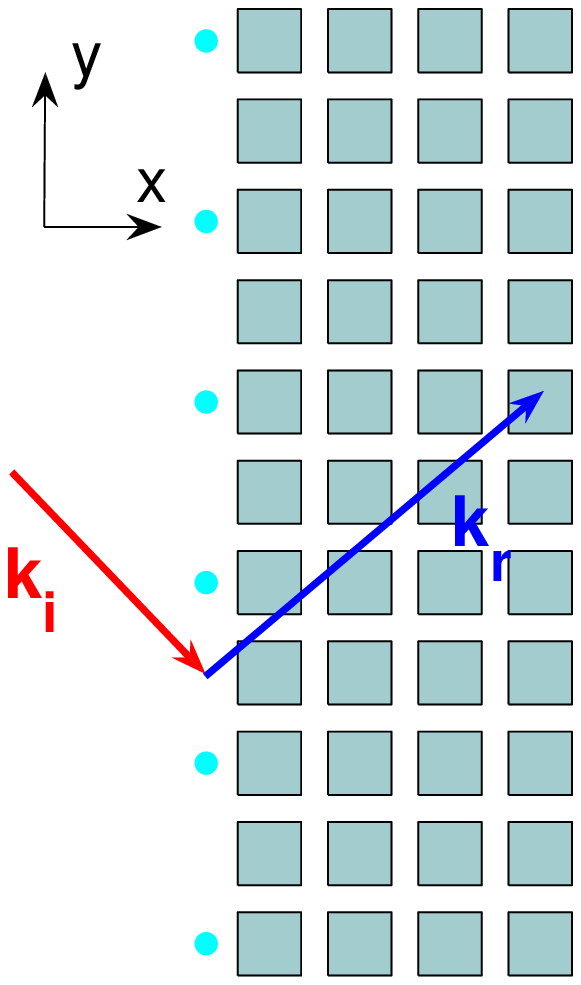}}
\caption{(Color online) Mechanism for NR and AANR for a square lattice PC with surface grating $a_s=2a$.
The red curve is the EFS in the PC and the green one is that in the air for $\o=0.219 \times 2\pi c/a$.
With the momentum kick $\Delta k=\pi/a$, an incident plane wave with ${\vec k}_i$ will be refracted 
 negatively into a plane wave with ${\vec k}_r$.}
\label{fig2}
\end{figure}

Consider a slab of this PC with
surface normal along the $\Gamma X$ direction. If one increases the
frequency, $\omega >\omega _{X}$, there will be a partial band gap for waves
incident to the air-PC interface since the Bloch waves have
$k_y\geq k_a$ and the incident plane wave with $k_y< k_a$
will be completely reflected. Here $k_a$ is the $k_y$ value of the crossing point of the EFS 
with the $XM$ boundary of the first Brillouin zone, as can be seen in Fig. \ref{fig1}
and Fig. \ref{fig2}.
For certain frequencies $\omega _{l}<\omega
<\omega _{M}$ with $\omega _{l}=0.2089\times 2\pi c/a$ when $k_a=\o/c$,
a flat slab of such PC is an omnidirectional reflector \cite{Fink}. 
For example for $\omega=0.219\times 2\pi c/a$ as shown in Fig. \ref{fig2}, 
there will be total external reflection for any incident plane wave. 
However for these frequencies, a surface grating with period
\be
a_{s}=2a
\ee
will give a momentum boost along the surface to the incident plane 
wave with incident angle $\t$ such that it will
be coupled to the Bloch waves inside the PC with transverse momentum 
$k_y=\pi/a+(\o/c)\sin\t$ if $\t$ is negative and
$k_y=-\pi/a+(\o/c)\sin\t$ if $\t$ is positive. The refracted wave will propagate 
on the opposite side of the surface normal with respect to the incident beam. 
Thus NR is achieved. This is illustrated in Fig. \ref{fig2}.
The effect of this surface grating is equivalent to bringing down the EFS 
around the $M$ point to the $X$ point for $\omega _{l}<\omega <\omega _{M}$. 
As we pointed out in Ref. \cite{Lu07},
it is the surface periodicity which determines the size of the EFS and 
the folding of the band structure.
Furthermore, if $\pi/a-k_a\geq \o/c$, AANR can be achieved. The upper limit for AANR is $\omega
_{u}=0.2192\times 2\pi c/a$. Thus we obtained a 4.7\% AANR around $\omega
_{u}$. 

\begin{figure}[htbp]
\vskip -2mm
\center{\includegraphics [angle=0,width=6.5cm]{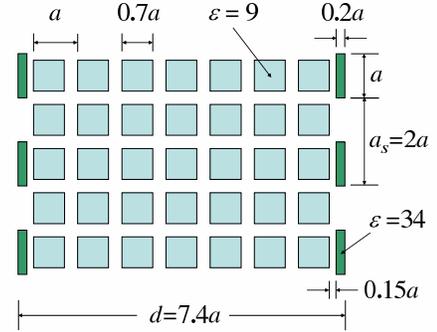}}
\vskip -2mm
\caption{(Color online) Details of slab made of a square lattice PC with surface grating $a_s=2a$.
For simplicity, the thickness $d$ of the slab is defined as the distance from the first surface to the 
last surface of the structure.}
\label{fig3}
\end{figure}

\begin{figure}[htbp]
\vskip -2mm
\center{\includegraphics [angle=0,width=8cm]{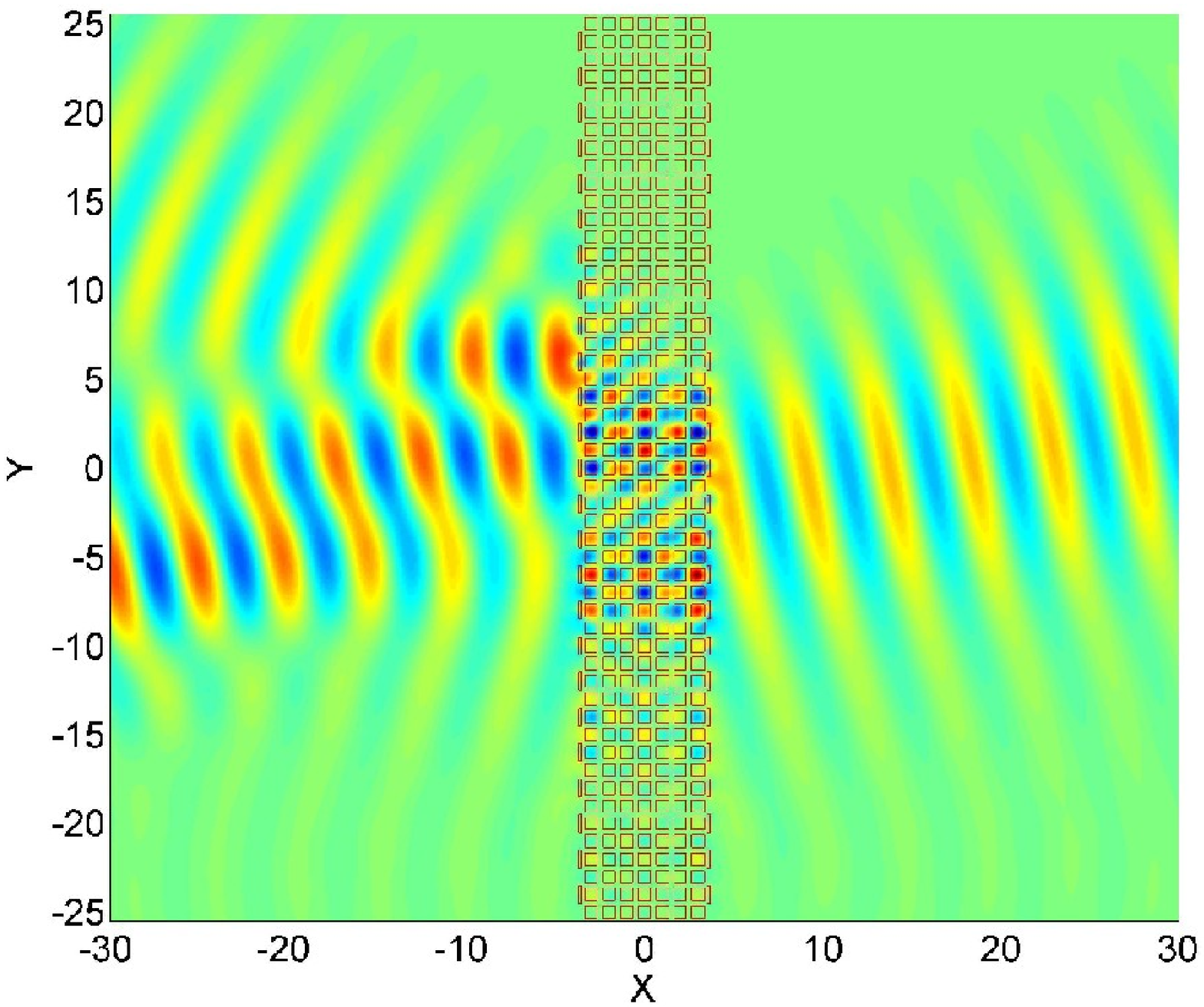}
\includegraphics [angle=0,width=8cm]{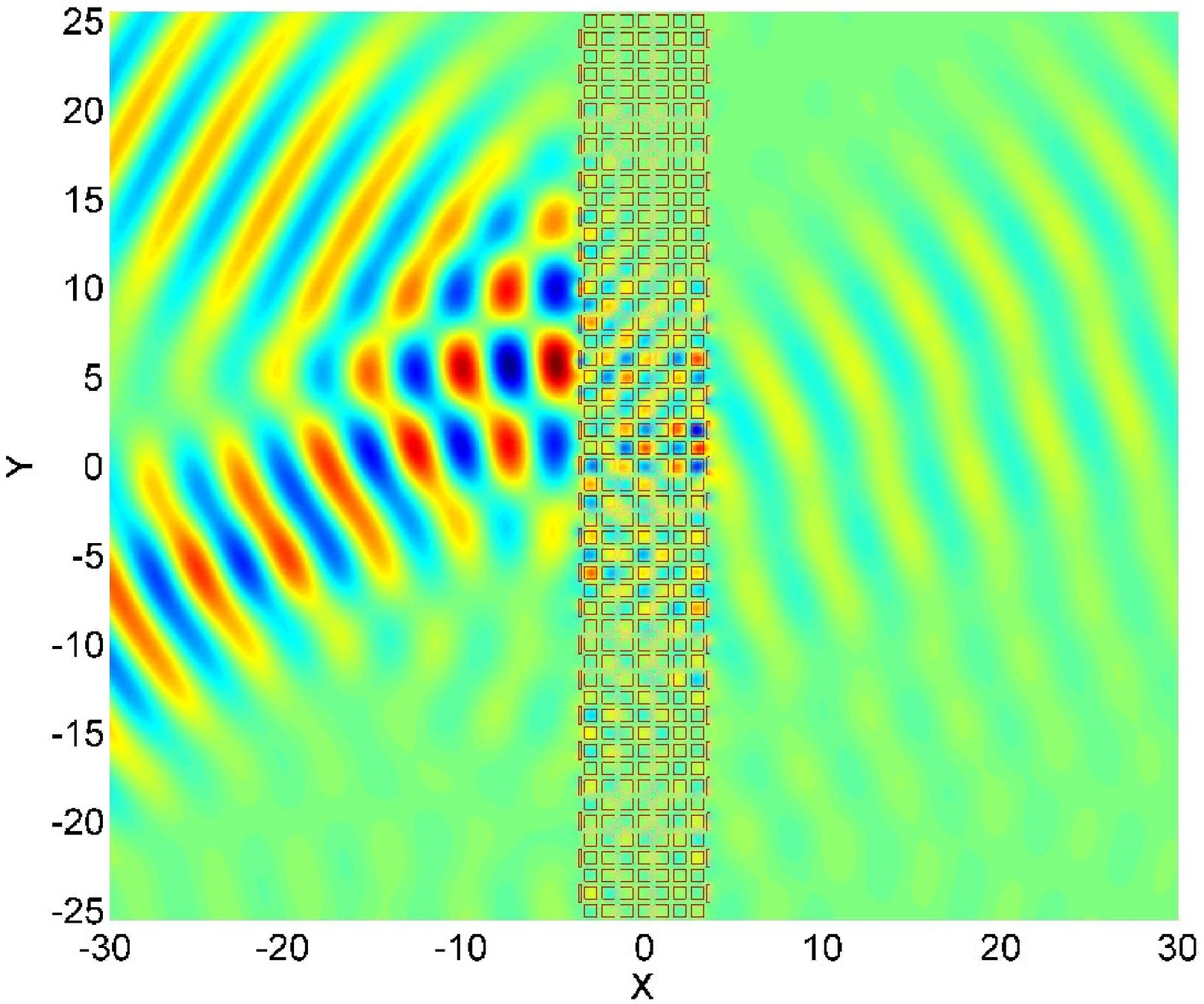}}
\vskip -2mm
\caption{(Color online) Negative lateral shift by a slab of PC given in Fig. \ref{fig3} 
for incident Gaussian beam with incident angles $15^\circ$ (top)
and $30^\circ$ (bottom) at $\o=0.219 \times 2\pi c/a$. The distance is measured in the unit of
lattice spacing $a$.}
\label{fig4}
\end{figure}

The above approach to NR and AANR is confirmed in our numeric simulations using FDTD \cite{Taflove}. 
Here we consider the lateral shift of an incident beam by a slab made of a square lattice PC. 
The detail of the slab is shown in Fig. \ref{fig3}. 
Negative lateral shifts are observed for different incident angles as shown in Fig. \ref{fig4} 
for beams at $\omega=0.219\times 2\pi c/a$. It can be verified that for this slab, 
AANR can be achieved for $0.2089\leq \o a/2\pi c\leq 0.2192$.
Note that the details of the surface grating is not essential except its period $a_s=2a$.
The grating can be holographic or an array of circular rods as long as it is not too thick.
For the specific surface grating shown in Fig. \ref{fig3}, 
the energy transmissions are $99.2\%$ and $4.7\%$ for
plane waves with incident angles $15^\circ$ and $30^\circ$, respectively.

\section{Flat lens with large $\protect\sigma $}

\begin{figure}[htbp]
\center{\includegraphics [angle=0,width=6.5cm]{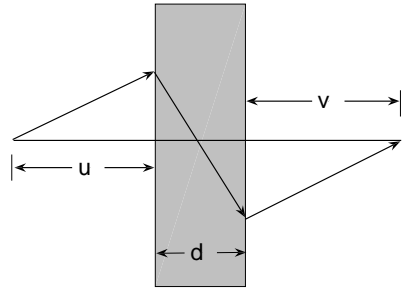}}
\vskip -6mm
\caption{A flat lens with lens equation $u+v=\s d$.}
\label{fig5}
\end{figure}

One prominent application of negative refraction is the Veselago-Pendry
perfect lens \cite{Pendry}. A flat slab of thickness $d$ can focus an object with distance 
$u$ on one side to a distance $v$ on the other side with 
$u+v=d$ if the refractive index $n=-1$. For a generalized flat lens 
without optical axis \cite{Lu05}, 
the lens equation takes the form
\begin{equation}
u+v=\sigma d
\end{equation}%
with $\sigma $ a material property, depending on the dispersion
characteristics of the flat lens. This is illustrated in Fig. \ref{fig5}.
This lens equation requires the following
form of the EFS at the operating frequency
\begin{equation}
k_{rx}=\kappa -\sigma \sqrt{\omega ^{2}/c^{2}-k_{y}^{2}}. \label{EFS-lens}
\label{EFS-flatlens}
\end{equation}%
The lens surface is in the $y$-direction. The surface normal is along the $x$-axis.
Here $k_{rx}$ is the longitudinal component of
the wave vector in the lens medium and $\kappa$ is the center of the EFS ellipse \cite{Lu05}.

Even though AANR can be realized in the first band along the $\Gamma M$
direction for a square lattice PC \cite{Luo02}, the EFS is very flat around
the lens normal. Thus one has $\sigma \ll 1$ \cite{Lu05}. 
Though $\sigma \sim 1$ has been reported in PCs with other structures \cite{ZhangXD,Gajic},
the focusing is still limited to the vicinity of the lens surface \cite{LiLin03}. 
For practical applications, we need large $\sigma $ so that the object and image can be
far away from the lens. We will show that with our new mechanism for AANR,
large $\sigma $ can be achieved.

\begin{figure}[htbp]
%\vskip 5mm
\center{\includegraphics [angle=0,width=7.5cm]{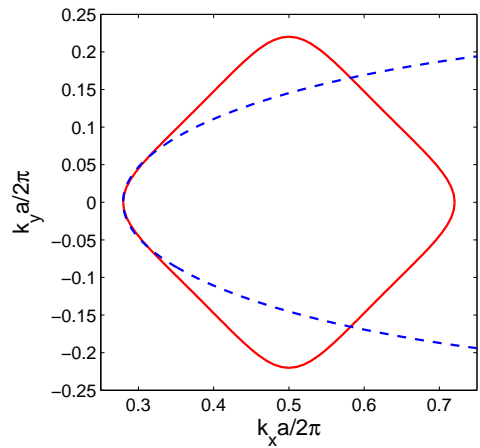}}
\caption{(Color online) Fitting of EFS of the TM modes at $\o a/2\pi c=0.219$ by Eq. (\ref{EFS-lens}).
Here $\s_0=4$ and $\kappa a/2\pi c=1.16$. Note that the center of EFS is shifted from
the $M$ point to the $X$ point due to the surface grating $a_s=2a$.}
\label{fig6}
\end{figure}

As we have stated in the previous section, the first Brillouin zone 
of a square lattice PC with surface grating $a_s=2a$
takes the shape of a rectangle instead of a square 
and its vertical size is reduced to $-\pi/a\leq k_y\leq \pi/a$.
The center of the EFS for $\omega _{l}\leq \omega \leq
\omega _{M}$ of the original PC is moved from the $M$ point to the $X$ point.
The fitting of the modified EFS for these frequency by Eq. (\ref{EFS-flatlens})
will give the lens property $\s$. An inspection of the band structure reveals that
the EFS is not elliptical. This results in incident angle dependent $\s$ \cite{Lu05}.
Nevertheless, the EFS can be can be fitted well with a constant $\s_0$ for small $k_y$
as shown in Fig. \ref{fig6}.
For the square lattice PC we have designed (Fig. \ref{fig3}), one has 
$\sigma_0 \sim 4$ for $\omega _{l}\leq \omega \leq
\omega _{u}$.

\begin{figure}[htbp]
\vskip -2mm
\center{\includegraphics [angle=0,width=8cm]{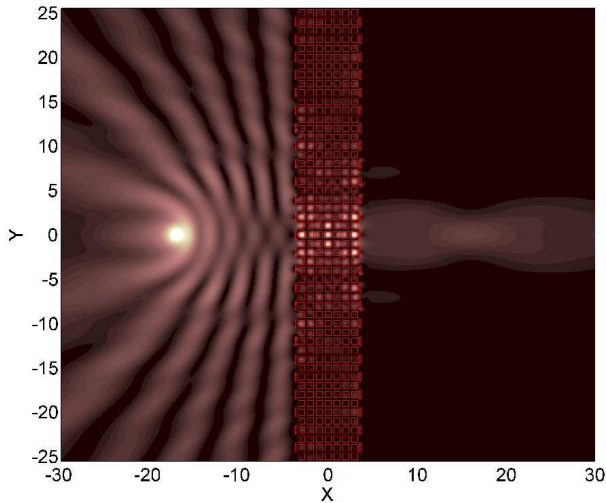}}
\vskip -2mm
\caption{(Color online) FDTD simulation of flat lens focusing of a point source. For better contrast effect, 
the field intensity at the point source is suppressed. The detail of the lens is given in Fig. \ref{fig3}.}
\label{fig7}
\end{figure}

Focusing by such a flat lens is shown in Fig. \ref{fig7}. 
For a point source with $u=13.6 a$, a clear focused image is obtained at $v=12.4 a$
for the operating frequency $\o=0.219 \times 2\pi c/a$, which is consistent with
the lens equation $u+v=\s_{\rm eff} d$ with $\s_{\rm eff}=3.5$ and $d=7.4 a$. 
There are two reasons for $\s_{\rm eff}<\s_0$. First that the EFS is elliptical
only for small $k_y=(\o/c)\sin\t$ and $\s\equiv-dk_{rx}/dk_x$ 
decreases with increasing incident angle $\t$. The effective $\s_{\rm eff}$ is an 
average and thus smaller than $\s_0$.
Second, the thickness of a PC slab is not a well-defined quantity. 
Here we simply define the lens 
thickness as the distance from the first surface to the last surface as shown in Fig. \ref{fig3}.
This may overestimate the effective thickness of the lens.

\begin{figure}[htbp]
\vskip -2mm
\center{\includegraphics [angle=0,width=7.5cm]{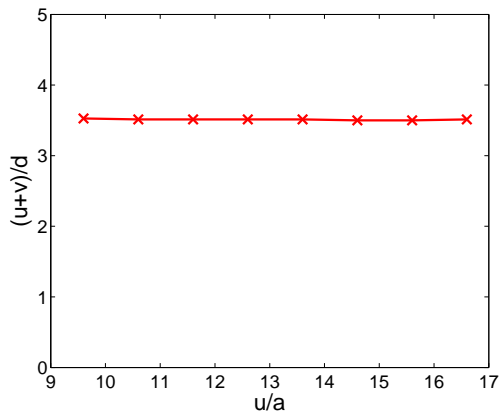}}
\vskip -2mm
\caption{The ratio of the object-image distance to the slab thickness $(u+v)/d$ vs. 
the object distance $u$ for the flat lens shown in Fig. \ref{fig3} at the operating frequency
$\o=0.219 \times 2\pi c/a$.}
\label{fig8}
\end{figure}

To further check the performance of this flat lens, we vary the object distance. 
In Fig. \ref{fig8} we show that the
ratio $(u+v)/d$ is almost constant and very close to $\s_0$ for different object distance $u$. 

\begin{figure}[htbp]
\vskip -2mm
\center{\includegraphics [angle=0,width=7.5cm]{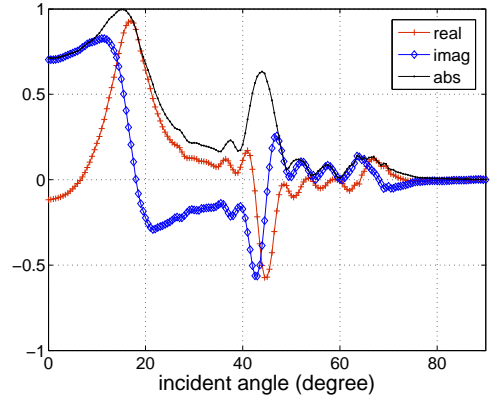}}
\vskip -2mm
\caption{(Color online) The transmission coefficient for plane wave incident on 
the flat lens shown in Fig. \ref{fig3}.}
\label{fig9}
\end{figure}

The primary concern of practical application is the power transmission
through the lens. As expected, the transmission through the flat lens is low
due to the impedance mismatch. However since the details of the grating on the
PC will not alter the scenario for NR, the power transmission can be
enhanced through careful engineering of the grating. In our simulation, we
find that the grating on the 2D PC with large dielectric constant has
strong power transmission. The parameters of the surface grating shown in Fig. \ref{fig3}
are not optimized. Further improvement of transmission may be possible.
For the grating parameters given in Fig. \ref{fig3}, the transmission 
coefficient is calculated and plotted in Fig. \ref{fig9}.

\section{Discussion and Conclusion}

In this paper, we have achieved NR using a new approach: photonic band gap
with surface grating. This approach to NR gives new window for AANR in 2D
PCs. This approach also gives flat lens made of these PCs with a large
object-image distance. Thus these flat lenses are able to image large and
far away objects.

Our new approach gives much larger window of AANR than previous realized. 
For example for a square lattice air holes in $\varepsilon =12$ with $r/a=0.35$
studied in Ref. \cite{Luo02}, our approach gives a lower limit $\omega _{l}=0.183\times 2\pi c/a
$ and an upper limit $\omega _{u}=0.206\times 2\pi c/a$, hence a fraction of
AANR frequency range of 11\% around $0.206\times 2\pi c/a$.
This range is much larger than the AANR range along the $\Gamma M$ direction for the TE modes.
This window of AANR is easier to locate.
The two limits are obtained from the crossing of the band with the light
lines around $X$ point and $M$ point, respectively. For the determination of the lower
limit $\o_l$, there is no need to compute the frequency
at which the radius of curvature of the contours along $\G M$
diverges \cite{Luo02}.

Our new approach to AANR can also be extended to three-dimensional photonic crystals.

\begin{acknowledgments}
This work was supported by the Air Force Research Laboratories, Hanscom
through FA8718-06-C-0045 and the National Science Foundation
through PHY-0457002.
\end{acknowledgments}

\end{document}